\documentclass[conference, colorlinks=true, linkcolor=black, citecolor=black, urlcolor=black, breaklinks=true]{IEEEtran}

\usepackage{amsmath,amssymb,amsfonts}

\usepackage{xcolor}
\usepackage{graphicx} 
\usepackage{hyperref} 
\usepackage{algorithm}

\begin{document}

\title{A Fully Hardware Implemented Accelerator Design in ReRAM Analog Computing without ADCs
\vspace{-10pt} 
}
\author{\IEEEauthorblockN{$\mathrm{Peng\ Dang}^{1,2,3},\ \mathrm{Huawei\ Li}^{1,3},\   \mathrm{Wei\ Wang}^{2}$}
\IEEEauthorblockA{$\mathrm{{}^{1} \textit{SKLP,} \ \textit{Institute} \ \textit{of} \ \textit{Computing} \ \textit{Technology,} \ \textit{Chinese} \ \textit{Academy} \ \textit{of} \ \textit{Sciences,} \ \textit{Beijing,} \ \textit{China}}$} 
\textit{$\mathrm{{}^{2} \textit{Peng} \ \textit{Cheng} \ \textit{Laboratory,} \ \textit{Shenzhen,} \ \textit{China,}}$}\
\textit{$\mathrm{{}^{3} \textit{University} \ \textit{of} \ \textit{Chinese} \ \textit{Academy} \ \textit{of} \ \textit{Sciences,} \ \textit{Beijing,} \ \textit{China}}$}\\
\textit{$\mathrm{dangpeng21@mails.ucas.ac.cn,\ lihuawei@ict.ac.cn,\ wangwei@pcl.ac.cn}$}\\}
\IEEEaftertitletext{\vspace{-15pt}}

\maketitle
\vspace{-20pt} 
\begin{abstract}
Emerging ReRAM-based accelerators process neural networks via analog Computing-in-Memory (CiM) for ultra-high energy efficiency. However, significant overhead in peripheral circuits and complex nonlinear activation modes constrain system energy efficiency improvements. This work explores the hardware implementation of the Sigmoid and SoftMax activation functions of neural networks with stochastically binarized neurons by utilizing sampled noise signals from ReRAM devices to achieve a stochastic effect. We propose a complete ReRAM-based Analog Computing Accelerator (RACA) that accelerates neural network computation by leveraging stochastically binarized neurons in combination with ReRAM crossbars. The novel circuit design removes significant sources of energy/area efficiency degradation, i.e., the Digital-to-Analog and Analog-to-Digital Converters (DACs and ADCs) as well as the components to explicitly calculate the activation functions. Experimental results show that our proposed design outperforms traditional architectures across all overall performance metrics without compromising inference accuracy.
\end{abstract}

\begin{IEEEkeywords}
ReRAM crossbar, Computing-in-Memory, Activation function, Stochastic binarization, Noise signal
\end{IEEEkeywords}

\section{I\textsc{ntroduction}}
With the rapid advancement of artificial intelligence (AI), Deep Neural Networks (DNNs) have been widely applied in fields such as image recognition and natural language processing  \cite{ambrogio2018equivalent}. However, traditional computing platforms are constrained by the "memory wall" bottleneck \cite{wan2022compute}, facing significant energy consumption challenges when processing large-scale neural networks. To address this issue, researchers have proposed CiM, a computing paradigm that integrates computation units directly within non-volatile memory, allowing for in-situ computation without data movement and thereby significantly reducing the energy overhead associated with data transfer. Among various emerging memory technologies, ReRAM has become the preferred device for accelerating DNN computations due to its low power consumption, high-speed operations, high endurance, and strong compatibility with CMOS technology \cite{aguirre2024hardware}.

\begin{figure}[!ht]
    \centering
    \vspace{-9pt}
    \includegraphics[width=0.93\linewidth]{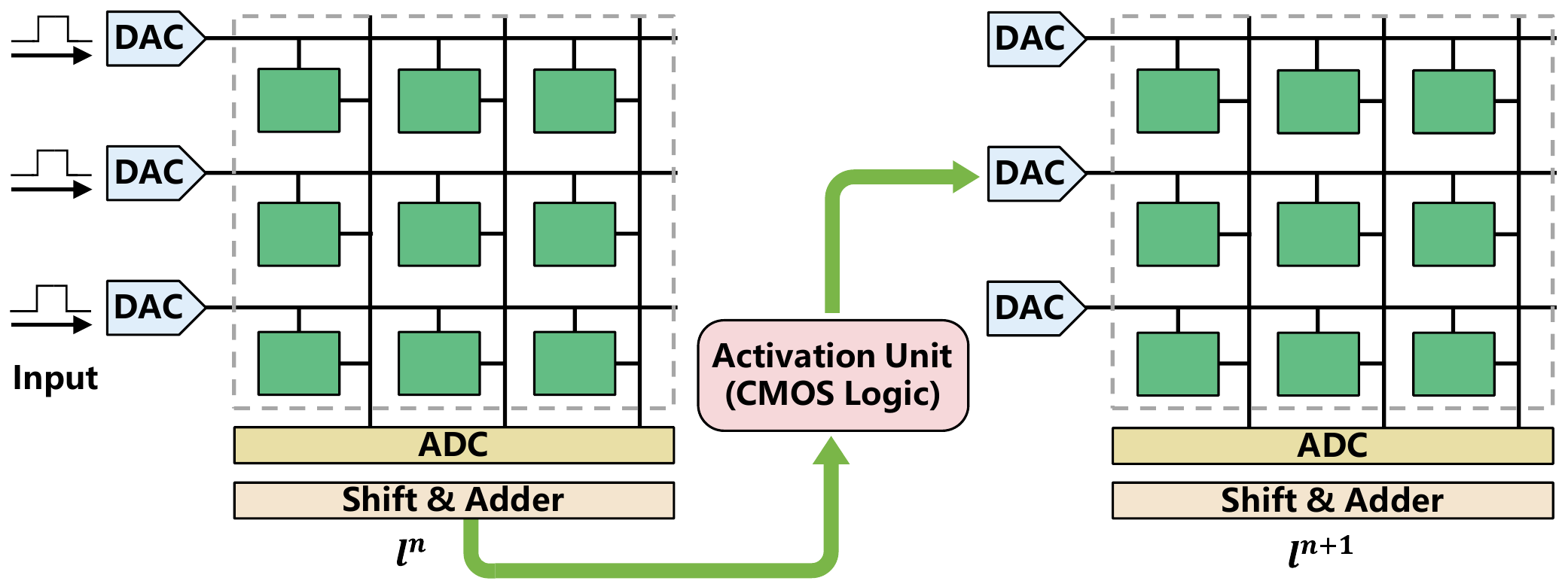}
    \caption{The conventional CiM architecture.}
    \vspace{-8pt} 
    \label{fig:main1} 
\end{figure}

In traditional ReRAM accelerators, as shown in Fig. \ref{fig:main1}, the multiply-accumulate (MAC) operations in neural networks are performed on crossbar arrays, while nonlinear activations are executed in digital CMOS logic units \cite{deutschmann2022towards,zhang2019design}. During a single computation cycle, the input vector is converted to analog voltages through DACs and applied to the crossbar array, where MAC operations are carried out based on Ohm's and Kirchhoff's laws \cite{guo2023fpcim,chi2016prime,chen2017neurosim+}. The results are then read via ADCs, and nonlinear activations are applied in the digital domain. In this process, DACs and ADCs serve as signal conversion bridges between the digital and analog domains. However, data from previous prototype verification systems indicate that DACs and ADCs account for up to 72\% of the total energy consumption and occupy as much as 81\% of the area \cite{ali2022compute}, with similar figures reported in other studies \cite{yao2020fully,lai2023enabling}. This suggests that optimizing ADCs or rethinking the hardware architecture is crucial to reducing energy consumption.

To address the above challenges, we propose a novel accelerator architecture, RACA, and its corresponding hardware implementation. RACA consists of ReRAM crossbar arrays and hardware-implemented Sigmoid and SoftMax activation circuits. The Sigmoid function utilizes noise current from ReRAM devices as a random number generator, while each layer’s output is activated through stochastic binarization. The SoftMax function follows a Winner-Takes-All (WTA) \cite{makhzani2015winner} rule, deriving classification results based on the cumulative probability distribution of multiple trials per neuron. Acting as simplified and efficient readout circuits, these activation circuits eliminate the need for costly DACs, ADCs, and explicit Sigmoid and SoftMax calculations, significantly reducing hardware complexity. Experimental results show that this design approach does not compromise inference accuracy.

\section{B\textsc{ackground}}
\subsection{Hardware Signal-to-Noise Ratio}
The noise current in electronic devices is an internal random variation present widely in electronic components, including ReRAM. Theoretically, noise can stem from various sources, including thermal noise, 
flicker noise and others. Thermal noise is a major contributor, arising from the random motion of electrons within the device and is directly proportional to temperature. Thermal noise can be estimated using the Nyquist formula \cite{johnson1928thermal}, which is expressed as:
\begin{equation}
i_{\text{RMS}} = \sqrt{4kTG \Delta f}
\label{eq:equation1}
\vspace{-3pt}
\end{equation}
where  $i_{\text{RMS}}$  is the root mean square (RMS) value of the noise current, $K$ is the Boltzmann constant, $T$ is the temperature, $G$ is the conductance, and $\Delta f$ is the bandwidth \cite{johnson1928thermal}. Please note that this is a simplified model and does not consider other potential sources of noise. 

Signal-to-Noise Ratio (SNR) is a metric used to measure the relative strength between the signal and noise, typically expressed in decibels (dB) \cite{gonugondla2020swipe}. The definition of SNR is the ratio of signal power to noise power, expressed as:
\begin{equation}
\mathrm{SNR}=10\cdot\log_{10}{\left(\frac{P_{\mathrm{signal}}}{P_{\mathrm{noise}}}\right)}
\end{equation}
where $P_{\mathrm{signal}}$ is the signal power, and $P_{\mathrm{noise}}$ is the noise power. In electronic circuits, noise current is typically related to noise power. Noise power can be expressed in terms of the RMS value of noise current and resistance, for example:
\begin{equation}
P_{\mathrm{noise}}=i_{\mathrm{RMS}}^2\cdot R
\end{equation}
where $i_{\mathrm{RMS}}$ is the RMS value of noise current, and $R$ is the resistance. 
This implies that the SNR depends on the signal power, the RMS value of the noise current, and the resistance. In practical circuit design, to enhance circuit performance, it is expected to reduce the noise current to improve the SNR.


\subsection{Weight Mapping}
A crossbar array of ReRAM allows for parallel scaling and computation of input signals $\left( x \right)$  and weight parameters $\left( W \right)$. Input signals are transmitted through the crossbar array's rows and columns of input lines. Each row corresponds to an input feature in the neural network, and each column corresponds to a neuron \cite{murmann2020mixed,gonugondla2020swipe,saxena2022towards}. During accelerated neural network computation, a voltage sequence is used to implement input signals, and the conductance of the crossbar array simulates artificial synapses to represent weights (Eq. \ref{eq:equation4}-\ref{eq:equation6}). In non-volatile memory like ReRAM, weight dynamic mapping can be achieved by adjusting the device's conductance \cite{giordano2019analog,sun2018fully}. Efficient dot product calculations are realized by applying physical laws. Subsequently, the current at the output represents the accumulated scaled signals (Eq. \ref{eq:equation7}).
\begin{equation}
G_0=\frac{G_{max}-G_{min}}{W_{max}-W_{min}}
\label{eq:equation4}
\end{equation}
\vspace{0.3pt}
\begin{equation}
G_{ref}=\frac{W_{max}\cdot G_{min}-W_{min}\cdot G_{max}}{W_{max}-W_{min}}
\end{equation}
\begin{equation}
V_j=x_j\cdot V_r
\label{eq:equation6}
\end{equation}
\begin{equation}
G_{ij}=W_{ij}\cdot G_0+G_{ref}
\label{eq:equation7}
\end{equation}
These equations demonstrate the weight mapping process, where ReRAM crossbar arrays perform parallelized scaling of input information, incorporating weights to accumulate and output results.

\section{M\textsc{ethodology}}
\subsection{Binary Stochastic Sigmoid Neurons}
\label{subsec:subsection1}
Stochastic Binary Neural Networks (SBNNs) use random thresholds to process neuron weights and activations, reducing computational and storage demands while maintaining model performance \cite{gonugondla2020swipe,she2019improving}. In SBNNs, the activation of each neuron is stochastically binarized during each forward pass \cite{darabi2018bnn+}. Binarization of neuron activation values using a stochastic threshold method, assuming $x$ is the original activation of the neuron, and $f\left(x\right)$ is the binarization function for activation, can be expressed as follows:
\begin{equation}
f\left( x \right) =\left\{ \begin{array}{l}
	1,\ x\ge p\\
	0,\ otherwise\\
\end{array} \right. 
\end{equation}
where $p$ is typically a stochastic threshold value between 0 and 1, determining the probability of activating binary values to 1 \cite{yan2019rram,tuma2016stochastic}. 
As the activation for SBNNs involves discrete binarization operations, this provides a basis for reforming CiM architectures that traditionally rely on ADCs and DACs for nonlinear activations.
\begin{figure}[!ht]
    \centering
    \vspace{-10pt}
    \includegraphics[width=0.9\linewidth]{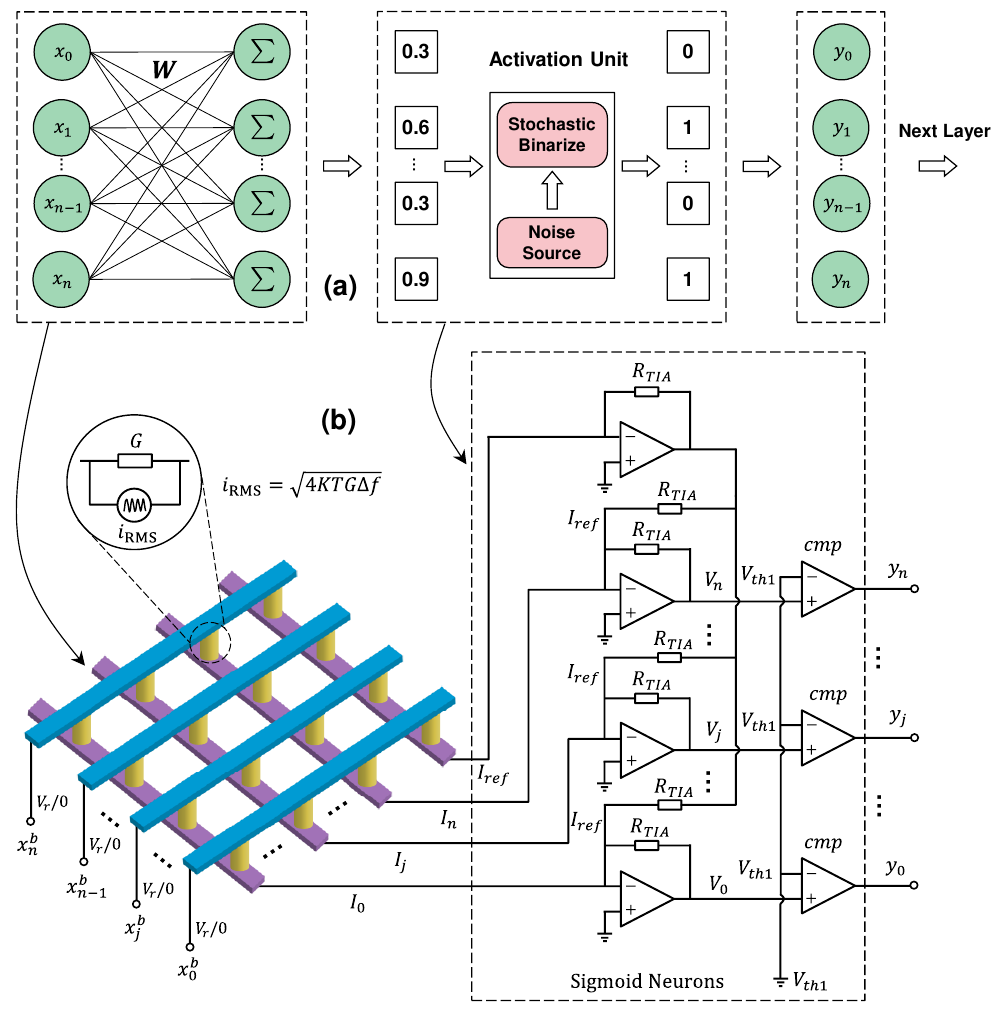}
    \caption{The implementation of the binary stochastic Sigmoid neurons by utilizing the Nyquist resistor noise. }
    \label{fig:main2} 
    \vspace{-10pt} 
\end{figure}

As shown in Fig. \ref{fig:main2}, we have constructed a highly simplified readout circuit for binary stochastic Sigmoid neurons. The output currents, termed Excitatory Postsynaptic Currents (EPSCs) \cite{williams1992simple,mahmoodi2019versatile}, are generated through the dot product computation of the input voltage with the conductance of the crossbar array. Subsequently, these EPSCs are converted into voltages via  Trans-impedance Amplifiers (TIAs). A simple subtraction circuit is employed to remove the reference voltage. Following this transformation, the resultant output voltage acts as a proportional analogue to the output of the neural network's software layer. This voltage is then compared using a voltage comparator to generate binary output. Due to the inherent Nyquist noise current of the ReRAM devices (as illustrated in Fig. \ref{fig:main2}(b) subplot), the output of the comparator is stochastically binarized When the signal intensity is appropriately reduced to fall within the range of noise current. Since the thermal noise current of each ReRAM device follows a Gaussian distribution, the activation probability for each neuron follows a Sigmoid shape (Eq. \ref{eq:IJ}-\ref{eq:Vth}).


{\small
\begin{equation}
I_j=\sum_{i}\left(V_i\cdot G_{ij}+I_{ij}^{\mathrm{noise}}\right)
\label{eq:IJ}
\end{equation}

\begin{equation}
I_{ref}=\sum_{i}\left(V_i\cdot G_{ref}+I_{ref}^{\mathrm{noise}}\right) 
\end{equation}
\begin{equation}
 I_{ij}^{\mathrm{noise}} \sim \mathcal{N}(0, 4kTG_{ij} \Delta f), \quad I_{ref}^{\mathrm{noise}} \sim \mathcal{N}(0, 4kTG_{ref} \Delta f) 
\end{equation}
\vspace{-10pt} 
\begin{align}
\overline{I_j}-\overline{I_{ref}} &= \sum_i V_i \cdot (G_{ij}-G_{ref}) \notag \\
&= V_r \cdot G_0 \cdot \sum_i W_{ij} \cdot x_i = V_r \cdot G_0 \cdot Z_j
\end{align}

\vspace{-12pt} 
\begin{align}
P\left( V_j > V_{th1} \right) &= P\left( I_j > I_{ref} \right) \notag \\[6pt]
&= \frac{1}{2}\left[ 1 + \text{erf}\left( \frac{I_j - I_{ref}}{\sqrt{4KT\Delta f \sum_i (G_{ij} + G_{ref})} \cdot \sqrt{2}} \right) \right] \notag \\[6pt]
&\approx \frac{1}{1 + e^{\frac{I_j - I_{ref}}{V_r \cdot G_0}}} = \frac{1}{1 + e^{-\sum_i W_{ij} \cdot x_i}}
\label{eq:Vth}
\end{align}
}

\subsection{WTA Binary Stochastic SoftMax Neuron}
\label{subsec:subsection2}
In the hardware implementation of the classifier, we utilize the WTA \cite{makhzani2015winner} strategy for handling multi-class classification tasks. This approach prioritizes the category with the highest score (or probability) as the decisive classification result. The WTA mechanism ensures that only one neuron is activated, with its activation intensity being mapped to a probability distribution. Ultimately, these individual probabilities are then amalgamated into a cumulative probability distribution, signifying the likelihood predicted to each category. We choose the category with the highest cumulative probability as the final predicted result during the classification process.
\begin{figure}[H]
    \centering
    \vspace{-12pt} 
    \includegraphics[width=0.9\linewidth]{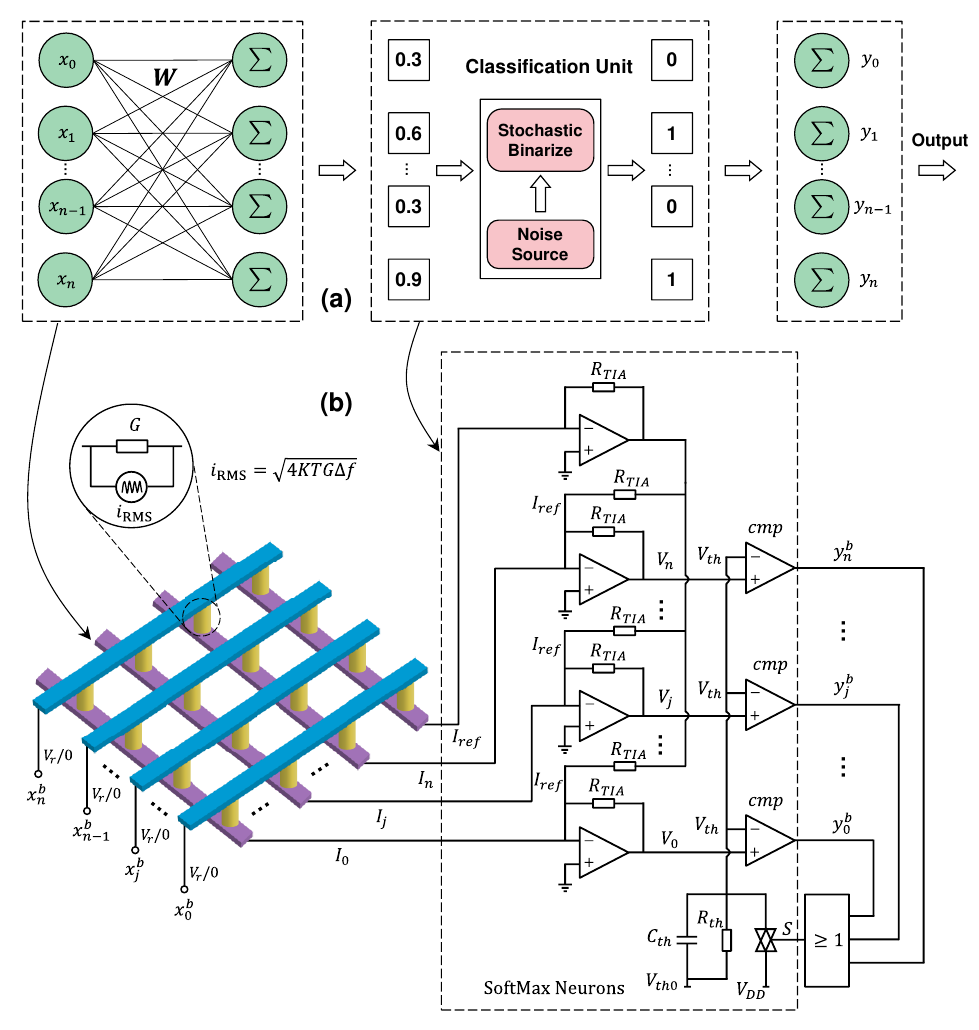}
    \caption{The implementation of the binary stochastic SoftMax neurons by utilizing the resistor noise and WTA rules. }
    \phantomsection
    \label{fig:main3} 
    \vspace{-12pt} 
\end{figure}

We compare the output voltage $V_j$ proportional to the adaptive threshold voltage $V_{th}$, where $\left(V_j \propto I_j - I_{\textit{ref}}\right)$, with the adaptive threshold voltage. The adaptive threshold, initially higher than the average of the output voltage at its static state $\left(V_{th0}\right)$, is pulled to the supply voltage when one of the neurons is activated (see Fig. \ref{fig:main3}). Subsequently, employing the WTA rule alongside the concept of cumulative probability distribution, the probability of activation for each neuron is given by the following formula:
\vspace{-5pt} 
\begin{equation}
P_{\text{WTA}}\left( y_{j}^{b}=1 \right) = \frac{P\left( y_{j}^{b}=1 \right)}{\sum_k{P}\left( y_{k}^{b}=1 \right)} \approx \frac{e^{\sum_i{w_{ij}x_i}}}{\sum_k{e}^{\sum_i{w_{ik}x_i}}}
\label{eq:equation14}
\end{equation}
This simulates the behavior of the SoftMax function. The Eq. \ref{eq:equation14} encapsulates the probabilistic framework within which each neuron's potential for activation is computed, mirroring the SoftMax function's role in normalizing neural network outputs to probabilistic distributions.

\subsection{RACA Architecture}
The proposed RACA architecture consists of cascaded layers of Sigmoid neurons (as detailed in Section \ref{subsec:subsection1}) and SoftMax neurons (as explained in Section \ref{subsec:subsection2}), where each layer integrates ReRAM crossbar arrays and activation circuits. Similar to the original SBNN, a DAC is used at the input stage to preserve the integrity of input data features. However, with our implemented Sigmoid and SoftMax neuron activation circuits, DACs and ADCs can be removed from the hidden and output layers. It should be noted that the hardware for computing the cumulative probability distribution of the classifier is not provided here, but it can be easily implemented at the output end with a simple counter. This mixed-signal design, which operates without ADCs and DACs, is also compatible with mainstream digital devices in terms of signal domain. Moreover, the number of neural network layers and specifications supported by this architecture can be flexibly configured by the user to accommodate various computational tasks.

\section{E\textsc{valuation}}
\subsection{Sigmoid Experimental Results}
\begin{figure}[!ht]
    \centering
    \vspace{-12pt} 
    \includegraphics[width=1\linewidth]{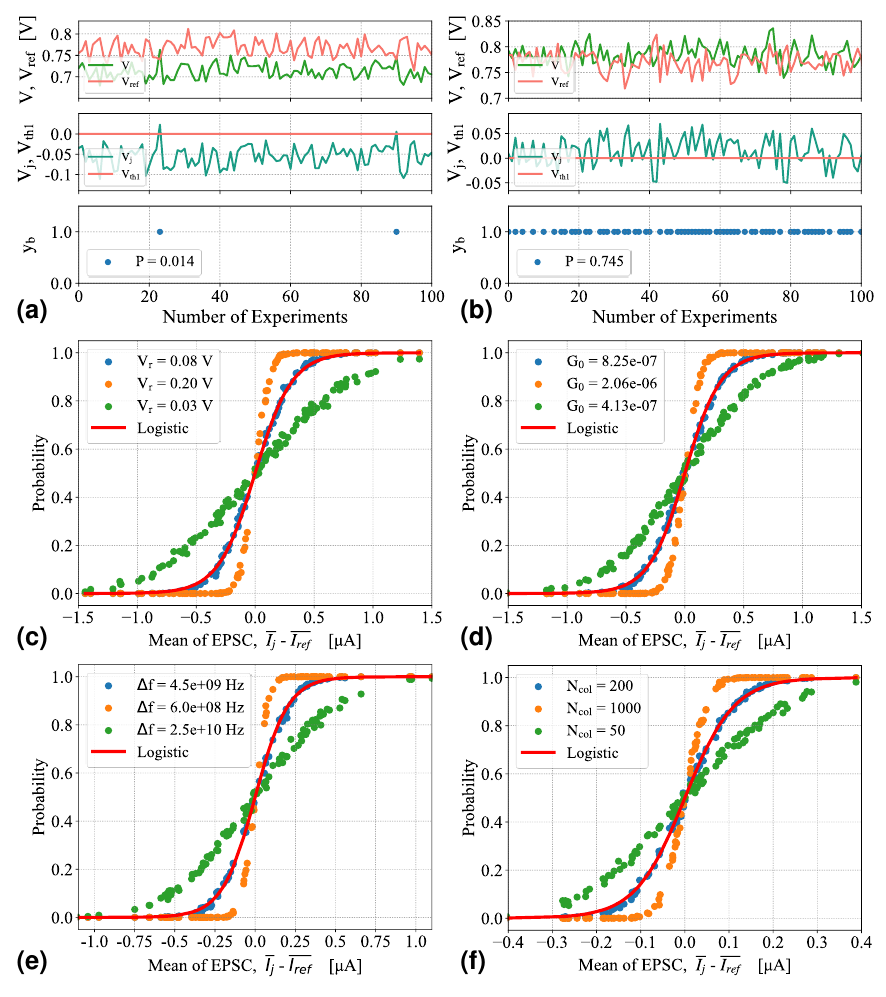}
    \vspace{-15pt} 
    \caption{Simulation results of Sigmoid neurons.}
    \label{fig:main5} 
    \vspace{-8pt} 
\end{figure}

Fig. \ref{fig:main5}(a) and (b) depict examples where the activation probabilities for two neurons are 0.014 and 0.745, respectively. The SNR can be adjusted by several factors: the amplitude $\left(V_r\right)$ of the input voltage as shown in Fig. \ref{fig:main5}(c), the scaling factor $\left(G_0\right)$ from algorithmic weight to conductance as seen in Fig. \ref{fig:main5}(d), the bandwidth $\left(\Delta f\right)$ of the readout circuit illustrated in Fig. \ref{fig:main5}(e), and the number $\left(N_{col}\right)$ of ReRAM devices in each column depicted in Fig. \ref{fig:main5}(f).The probability of neuron activation can closely emulate the logistic function, a widely used Sigmoid activation function in deep neural networks. Notably, the statistical probabilities displayed in Fig. \ref{fig:main5}(c)-(f), obtained through continuous sampling experiments, are solely used for analysis and comparison. 

\subsection{SoftMax Experimental Results}
\begin{figure}[!ht]
    \centering
    \vspace{-15pt}
    \includegraphics[width=0.95\linewidth]{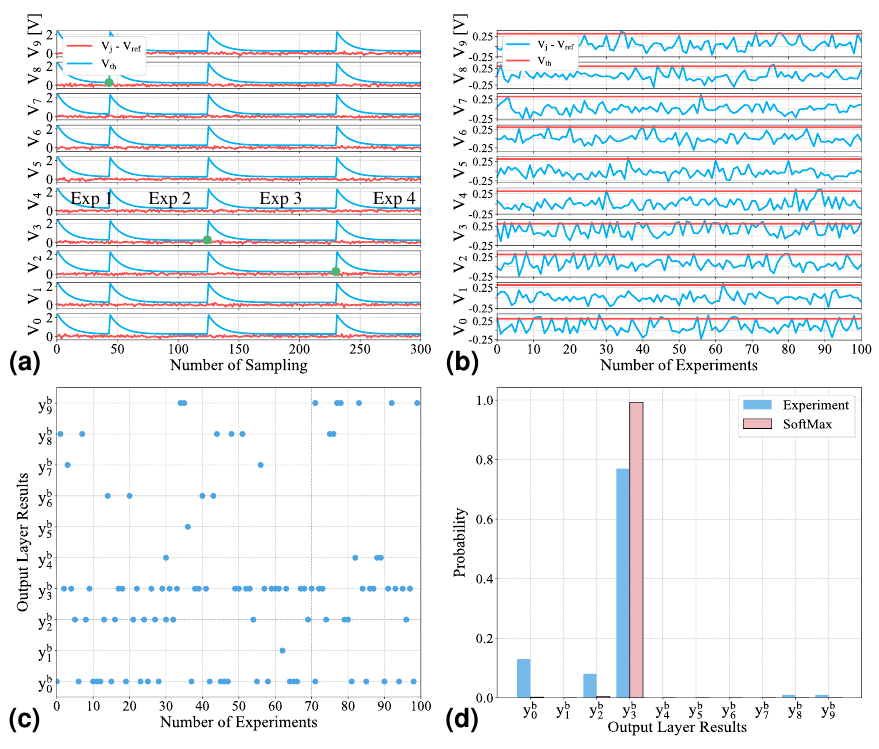}
    \vspace{-5pt}
    \caption{Simulation results of SoftMax neurons.}
    \label{fig:main6} 
    \vspace{-12pt}
\end{figure}

Fig. \ref{fig:main6}(a) presents illustrative traces of the output voltage over time, compared with the adaptive threshold voltage, for ten WTA binary stochastic neurons in three consecutive decision experiments. This indicates that only one neuron is activated in each decision experiment. Fig. \ref{fig:main6}(b) presents the output voltage compared to the threshold voltage for 100 decision experiments, and Fig. \ref{fig:main6}(c) displays the corresponding raster plot for the ten neurons. Fig. \ref{fig:main6}(d) shows the statistical probability of activated neurons compared with the experimental results in Fig. \ref{fig:main6}(c), along with the ideal SoftMax neuron's software-calculated results. The results suggest that both models attained the highest cumulative probability on the same neuron in the output layer, indicating identical predictions from both.

\subsection{RACA Accelerators Architecture Experimental Results}
The proposed design's inference performance was tested using a fully trained fully connected neural network (FCNN) with a structure [784, 500, 300, 10] on the MNIST dataset. Binary stochastic Sigmoid neurons were used for the first two layers, and WTA binary stochastic SoftMax neurons were used for the final classification layer. For a specific neural network architecture, $N_{col}$, $G_0$ are immutable constants. In this case, proper SNR can be achieved to mimic the desired Sigmoid function by adjusting the $V_r$ and $\Delta f$ parameters. In evaluation, we utilized Ag:Si devices fabricated with 32 nm process. To lower the SNR, the read voltage should be much smaller than the usual read voltage of ReRAM, resulting in lower energy consumption, in addition to the reduced complexity of the neuron circuits.

\begin{figure}[!ht]
    \centering
    \vspace{-8pt}
    \includegraphics[width=1\linewidth]{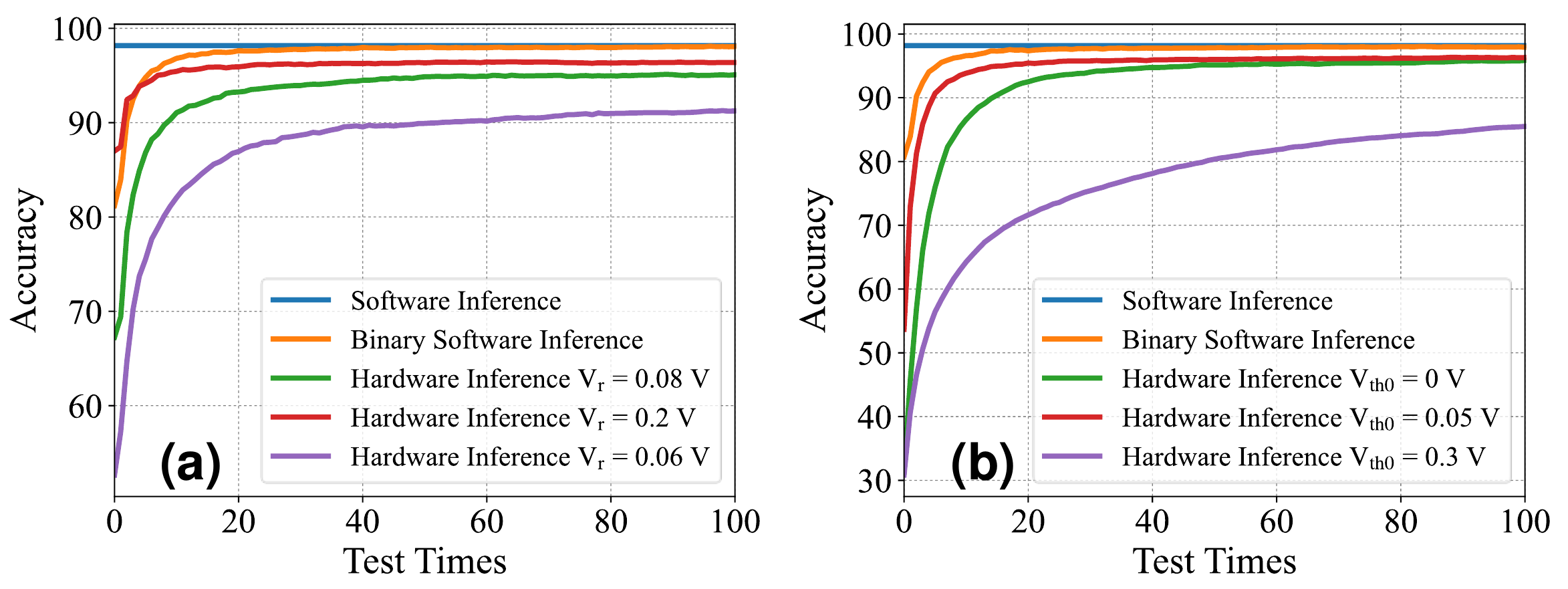}
    \vspace{-12pt}
    \caption{Experimental results of RACA. (a) Test results for varying the SNR in the Sigmoid layers. (b) Test results for varying the threshold voltage in the SoftMax layer.}
    \label{fig:main7} 
    \vspace{-10pt}
\end{figure}

\textbf{Test Accuracy:} Due to the stochastic nature of the inference, a single inference test meets an unavoidable decrease in recognition accuracy. However, repeating the stochastic inference and making a majority vote of the WTA SoftMax neurons could quickly improve the overall recognition accuracy, as shown in Fig.\ref{fig:main7}(a) and (b). The findings also suggest that there is a wide range of  values that can be utilized for our designed activation function, indicating improved robustness of the system. For the SoftMax neurons, the rest state threshold voltage  is an important parameter. Small  undermines the approximation in Eq. \ref{eq:equation1}, while high  decreases the activation probability of each neuron, which prolongs a single decision time. Our results demonstrate that with  set to 0.05V, the system's inference accuracy increases with the number of tests and achieves a final accuracy of 96.7\%. When  is set to 0 V, an accuracy of 96\% is achieved, although it falls slightly below the accuracy obtained with =0.05 V for fewer test iterations. 

\textbf{Performance Metrics:} We used a modified NeuroSim \cite{chen2017neurosim+} to estimate the hardware metrics of accelerators employing a 1-bit ADC and those adopting RACA, with the results presented in Table \ref{tab:Table2}. As shown in the table, replacing ADCs with comparators significantly enhances energy efficiency and reduces area overhead. Here, $\downarrow$ indicates the percentage decrease, and $\uparrow$ indicates the percentage increase.
\begin{table}[!ht]
    \centering
    \vspace{-3pt}  
    \setlength{\tabcolsep}{3pt}  
    \renewcommand{\arraystretch}{1.5} 
    \small  
    \begin{tabular}{|c|c|c|c|c|c|} \hline 
         Schemes &  \multicolumn{2}{c|}{1-bit ADC} &  \multicolumn{2}{c|}{RACA} & Change (\%)\\ \hline 
         Energy Consumption ($\times 10^5$ pJ) &  \multicolumn{2}{c|}{8.7} &  \multicolumn{2}{c|}{3.63} & $\downarrow$ 58.29 \\ \hline 
         Area Overhead ($\text{mm}^2$) &  \multicolumn{2}{c|}{8.51} &  \multicolumn{2}{c|}{5.24} & $\downarrow$ 38.43 \\ \hline 
         Energy Efficiency (TOPS/W) &  \multicolumn{2}{c|}{61.3} &  \multicolumn{2}{c|}{148.58} & $\uparrow$ 142.37 \\ \hline
    \end{tabular}
    \caption{Hardware metrics comparison for FCNN on MNIST.}
    \label{tab:Table2}
    \vspace{-10pt}
\end{table}

\section{C\textsc{onclusion}}
We propose a novel circuit design that leverages hardware noise signals to emulate the Sigmoid and SoftMax functions in neural network inference. These stochastically activated neurons are achieved by appropriately reducing the SNR of the crossbar array in artificial synapses. We circumvent the need for precisely sensing the output of the crossbar array and explicitly computing the activation functions, thereby realizing a highly simplified, efficient, and biologically inspired method for neural network inference.

\bibliographystyle{ieeetr}
\bibliography{references.bbl}

\begin{thebibliography}{10}

\bibitem{ambrogio2018equivalent}
S.~Ambrogio, P.~Narayanan, H.~Tsai, R.~M. Shelby, I.~Boybat, C.~Di~Nolfo,
  S.~Sidler, M.~Giordano, M.~Bodini, N.~C. Farinha, {\em et~al.},
  ``Equivalent-accuracy accelerated neural-network training using analogue
  memory,'' {\em Nature}, vol.~558, no.~7708, pp.~60--67, 2018.

\bibitem{wan2022compute}
W.~Wan, R.~Kubendran, C.~Schaefer, S.~B. Eryilmaz, W.~Zhang, D.~Wu, S.~Deiss,
  P.~Raina, H.~Qian, B.~Gao, {\em et~al.}, ``A compute-in-memory chip based on
  resistive random-access memory,'' {\em Nature}, vol.~608, no.~7923,
  pp.~504--512, 2022.

\bibitem{aguirre2024hardware}
F.~Aguirre, A.~Sebastian, M.~Le~Gallo, W.~Song, T.~Wang, J.~J. Yang, W.~Lu,
  M.-F. Chang, D.~Ielmini, Y.~Yang, {\em et~al.}, ``Hardware implementation of
  memristor-based artificial neural networks,'' {\em Nature communications},
  vol.~15, no.~1, p.~1974, 2024.

\bibitem{deutschmann2022towards}
L.~Deutschmann, J.~M{\"u}ller, M.~R. Fadiheh, D.~Stoffel, and W.~Kunz,
  ``Towards a formally verified hardware root-of-trust for data-oblivious
  computing,'' in {\em Proceedings of the 59th ACM/IEEE Design Automation
  Conference}, pp.~727--732, 2022.

\bibitem{zhang2019design}
W.~Zhang, X.~Peng, H.~Wu, B.~Gao, H.~He, Y.~Zhang, S.~Yu, and H.~Qian, ``Design
  guidelines of rram based neural-processing-unit: A joint
  device-circuit-algorithm analysis,'' in {\em Proceedings of the 56th Annual
  Design Automation Conference 2019}, pp.~1--6, 2019.

\bibitem{guo2023fpcim}
Y.-C. Guo, W.-T. Lin, T.-H. Hou, and T.-S. Chang, ``Fpcim: A fully-parallel
  robust reram cim processor for edge ai devices,'' in {\em 2023 IEEE
  International Symposium on Circuits and Systems (ISCAS)}, pp.~1--5, IEEE,
  2023.

\bibitem{chi2016prime}
P.~Chi, S.~Li, C.~Xu, T.~Zhang, J.~Zhao, Y.~Liu, Y.~Wang, and Y.~Xie, ``Prime:
  A novel processing-in-memory architecture for neural network computation in
  reram-based main memory,'' {\em ACM SIGARCH Computer Architecture News},
  vol.~44, no.~3, pp.~27--39, 2016.

\bibitem{chen2017neurosim+}
P.-Y. Chen, X.~Peng, and S.~Yu, ``Neurosim+: An integrated device-to-algorithm
  framework for benchmarking synaptic devices and array architectures,'' in
  {\em 2017 IEEE International Electron Devices Meeting (IEDM)}, pp.~6--1,
  IEEE, 2017.

\bibitem{ali2022compute}
M.~Ali, S.~Roy, U.~Saxena, T.~Sharma, A.~Raghunathan, and K.~Roy,
  ``Compute-in-memory technologies and architectures for deep learning
  workloads,'' {\em IEEE Transactions on Very Large Scale Integration (VLSI)
  Systems}, vol.~30, no.~11, pp.~1615--1630, 2022.

\bibitem{yao2020fully}
P.~Yao, H.~Wu, B.~Gao, J.~Tang, Q.~Zhang, W.~Zhang, J.~J. Yang, and H.~Qian,
  ``Fully hardware-implemented memristor convolutional neural network,'' {\em
  Nature}, vol.~577, no.~7792, pp.~641--646, 2020.

\bibitem{lai2023enabling}
Y.-S. Lai, S.-H. Chen, and Y.-H. Chang, ``Enabling highly-efficient dna
  sequence mapping via reram-based tcam,'' in {\em 2023 IEEE/ACM International
  Symposium on Low Power Electronics and Design (ISLPED)}, pp.~1--6, IEEE,
  2023.

\bibitem{makhzani2015winner}
A.~Makhzani and B.~J. Frey, ``Winner-take-all autoencoders,'' {\em Advances in
  neural information processing systems}, vol.~28, 2015.

\bibitem{johnson1928thermal}
J.~B. Johnson, ``Thermal agitation of electricity in conductors,'' {\em
  Physical review}, vol.~32, no.~1, p.~97, 1928.

\bibitem{gonugondla2020swipe}
S.~K. Gonugondla, A.~D. Patil, and N.~R. Shanbhag, ``Swipe: Enhancing
  robustness of reram crossbars for in-memory computing,'' in {\em Proceedings
  of the 39th International Conference on Computer-Aided Design}, pp.~1--9,
  2020.

\bibitem{murmann2020mixed}
B.~Murmann, ``Mixed-signal computing for deep neural network inference,'' {\em
  IEEE Transactions on Very Large Scale Integration (VLSI) Systems}, vol.~29,
  no.~1, pp.~3--13, 2020.

\bibitem{saxena2022towards}
U.~Saxena, I.~Chakraborty, and K.~Roy, ``Towards adc-less compute-in-memory
  accelerators for energy efficient deep learning,'' in {\em 2022 Design,
  Automation \& Test in Europe Conference \& Exhibition (DATE)}, pp.~624--627,
  IEEE, 2022.

\bibitem{giordano2019analog}
M.~Giordano, G.~Cristiano, K.~Ishibashi, S.~Ambrogio, H.~Tsai, G.~W. Burr, and
  P.~Narayanan, ``Analog-to-digital conversion with reconfigurable function
  mapping for neural networks activation function acceleration,'' {\em IEEE
  Journal on Emerging and Selected Topics in Circuits and Systems}, vol.~9,
  no.~2, pp.~367--376, 2019.

\bibitem{sun2018fully}
X.~Sun, X.~Peng, P.-Y. Chen, R.~Liu, J.-s. Seo, and S.~Yu, ``Fully parallel
  rram synaptic array for implementing binary neural network with (+ 1,- 1)
  weights and (+ 1, 0) neurons,'' in {\em 2018 23rd Asia and South Pacific
  Design Automation Conference (ASP-DAC)}, pp.~574--579, IEEE, 2018.

\bibitem{she2019improving}
X.~She, Y.~Long, and S.~Mukhopadhyay, ``Improving robustness of reram-based
  spiking neural network accelerator with stochastic
  spike-timing-dependent-plasticity,'' in {\em 2019 International Joint
  Conference on Neural Networks (IJCNN)}, pp.~1--8, IEEE, 2019.

\bibitem{darabi2018bnn+}
S.~Darabi, ``Regularized binary network training,'' {\em arXiv:1812.11800},
  2018.

\bibitem{yan2019rram}
B.~Yan, Q.~Yang, W.-H. Chen, K.-T. Chang, J.-W. Su, C.-H. Hsu, S.-H. Li, H.-Y.
  Lee, S.-S. Sheu, M.-S. Ho, {\em et~al.}, ``Rram-based spiking nonvolatile
  computing-in-memory processing engine with precision-configurable in situ
  nonlinear activation,'' in {\em 2019 Symposium on VLSI Technology},
  pp.~T86--T87, IEEE, 2019.

\bibitem{tuma2016stochastic}
T.~Tuma, A.~Pantazi, M.~Le~Gallo, A.~Sebastian, and E.~Eleftheriou,
  ``Stochastic phase-change neurons,'' {\em Nature nanotechnology}, vol.~11,
  no.~8, pp.~693--699, 2016.

\bibitem{williams1992simple}
R.~J. Williams, ``Simple statistical gradient-following algorithms for
  connectionist reinforcement learning,'' {\em Machine learning}, vol.~8,
  pp.~229--256, 1992.

\bibitem{mahmoodi2019versatile}
M.~Mahmoodi, M.~Prezioso, and D.~Strukov, ``Versatile stochastic dot product
  circuits based on nonvolatile memories for high performance neurocomputing
  and neurooptimization,'' {\em Nature communications}, vol.~10, no.~1,
  p.~5113, 2019.

\end{thebibliography}

\end{document}